# Chemical Bonding in Aqueous Ferrocyanide: Experimental and Theoretical X-ray Spectroscopic Study

*Nicholas Engel,[1] Sergey I. Bokarev,*[,2] Edlira Suljoti,[1] Raul Garcia-Diez,[1] Kathrin M. Lange,[1] Kaan Atak,[1] Ronny Golnak,[1] Alexander Kothe,[1] Marcus Dantz,[1] Oliver Kühn,[2] Emad F. Aziz*[,1]*

[1]Joint Ultrafast Dynamics Lab in Solutions and at Interfaces (JULiq)

at Helmholtz-Zentrum Berlin für Materialien und Energie, Albert-Einstein-Strasse 15,

12489 Berlin, Germany

and Freie Universität Berlin, Department of Physics, Arnimallee 14, 14195 Berlin, Germany.

[2]Institut für Physik, Universität Rostock, Universitätsplatz 3, 18055 Rostock, Germany.

**ABSTRACT**: Resonant inelastic X-ray scattering (RIXS) and X-ray absorption (XA) experiments at the iron L- and nitrogen K-edge are combined with high-level first principles restricted active space self-consistent field (RASSCF) calculations for a systematic investigation of the nature of the chemical bond in potassium ferrocyanide in aqueous solution. The atom- and site-specific RIXS excitations allow for direct observation of ligand-to-metal (Fe L-edge) and metal-to-ligand (N K-edge) charge transfer bands and thereby evidence for strong σ-donation and π-back-donation. The effects are identified by comparing experimental and simulated spectra related to both the unoccupied and occupied molecular orbitals in solution.

**INTRODUCTION**

Transition metals (TMs) play major roles throughout many fields in science. In metalloproteins, TMs are often key-centers for protein functions[1-2] and more than half of all proteins is estimated to be of that type.[3] Such complexes, e.g., enzymes in the human body, can play a role in inactivating carcinogenic substances.[4] Furthermore, TMs are found at the centers of industrial TM complex-catalysts.[5-10] Since catalytic reactions are fundamentally important, the necessity for understanding their functional mechanisms is self-evident. Catalytic processes are connected with chemical bond-reorganizations, and accordingly, understanding such mechanism is vital for the future development of catalysts. Detailed insight can be attained by the investigation of the electronic structure using X-ray absorption spectroscopy (XAS) and resonant inelastic X-ray scattering (RIXS) techniques.

XAS is an element-specific method since the atomic core electrons (with specific binding energies) are selectively excited to the unoccupied valence molecular orbitals (MOs). Thereby XAS probes the local, atom-specific unoccupied MOs in complex molecules consisting of several atoms. The resonant excitation of the core electron leaves the system in a highly excited state. Subsequently, the core orbital is refilled by an electron from the occupied valence MOs. Accordingly, RIXS probes atom-specific occupied valence MOs. All transition processes described here are sensitive to the MO symmetry by virtue of (dipole) selection rules.

Whereas both biological and industrial catalysis occurs mainly in solution, it has not been possible to use soft X-ray spectroscopies for the investigation of liquid phase spectra for a long time. Recent experimental development[11] has enabled the successful application of these techniques to the model catalyst $Fe(CO)_5$ in solution.[12] In the present study, we address another model TM complex, namely potassium ferrocyanide ($K_4[Fe(CN)_6]$),[13] in aqueous solution.

Ferrocyanide and its oxidized form, ferricyanide, have copious applications in industry and science. Here just a few of them are listed: pigments,[14] anticaking agents,[15] and redox reactions.[16-18] In particular, ferroycanide is a classic model system for understanding the chemical bonding in TM complexes. $[Fe(CN)_6]^{4-}$ is an octahedral complex with low-spin singlet ground state of $A_{1g}$ symmetry with electronic configuration $...(\sigma(e_g))^4(3d(t_{2g}))^6(\sigma^*(e_g))^0(\pi^*_{CN}(t_{2g}))^0....$[19] The spectroscopy is determined by the unoccupied $\sigma^*(e_g)$ and $\pi^*_{CN} + d_{Fe}(t_{2g})$ MOs.

For TM complexes containing ligands with π-bonds, the bond between metal and ligand is generally described as σ-donation and π-backdonation within the valence bond (VB) approach.[19-23] Thereby, the occupied ligand and unoccupied or partly occupied metal orbitals mix, leading to electron density redistribution from the ligand to the metal, called σ-donation. The reverse process, i.e., the mixing between the metal's occupied orbitals with the unoccupied ones of the ligand is referred to as π-backdonation. In the case of $[Fe(CN)_6]^{4-}$, specifically in σ-donation, charge is transferred from the highest occupied σ-type lone pair orbitals of cyanide (CN⁻) to the unfilled metal $\sigma^*(e_g)$ orbitals and in π-backdonation metal $3d(t_{2g})$ electrons are transferred into the lowest unoccupied antibonding 2π* ligand orbitals.[13]

In this manuscript, the strength of metal-ligand orbital mixing and the associated degree of σ-donation and π-backdonation are determined for the model catalyst aqueous ferrocyanide. The effects are quantified by applying *ab initio* RASSCF calculations to interpret the experimental Fe L-edge (metal center) and N K-edge (ligand) XAS and RIXS spectra. Furthermore, we compare the donation and backdonation properties in the $[Fe(CN)_6]^{4-}$ complex, as revealed in this study, with those of the $Fe(CO)_5$ catalyst.[12]

**EXPERIMENTAL SECTION**

0.4M solution was prepared by dissolving potassium hexacyanoferrate(II) trihydrate (purity >99%, Sigma-Aldrich) in Milli-Q water and subsequent filtration. To avoid photodissociation and oxidation of iron in the neutral pH solution, the sample was protected from light and was prepared freshly directly before each measurement.

Total fluorescence yield (TFY), partial fluorescence yield (PFY), and RIXS spectra were recorded at the L-edge of iron and the K-edge of nitrogen using the recently developed LiXEdrom set-up at BESSY II undulator beamline U41-PGM of Helmholtz-Zentrum Berlin. The PFY and RIXS spectra were measured with the X-ray emission spectrometer that operates in Rowland geometry.[11] The synchrotron beam was linearly polarized with polarization vector perpendicular to the liquid-beam. Using a spectrometer, the photons emitted in the plane of the electric field of the synchrotron beam were detected. The liquid beam of 0.4M aqueous ferrocyanide solution was introduced to the experimental chamber using the liquid micro-jet technique. A detailed description of the LiXEdrom setup and the micro-jet technique was presented previously.[11] The pressure in the interaction chamber was ca. $1.5 \times 10^{-5}$ mbar and the spectral resolving power $E/\Delta E$ of the beamline was larger than 2000 at both the N K-edge and Fe L-edge.

Notice, that for the prepared sample, the ferrocyanide anions are repelled from the surface.[24-25] Accordingly, the presented data can be attributed to bulk-solvated sample.

**COMPUTATIONAL DETAILS**

For the theoretical investigations the $[Fe(CN)_6]^{4-}$ ion was embedded in a polarizable continuum (PCM)[26] mimicking the influence of the water solvent. The equilibrium geometries for low-spin singlet ground state with $O_h$ point group symmetry were obtained at the

BLYP/LANL2DZ (Fe atom)[27] and 6-311+G(d) (C and N atoms)[28] level using the Gaussian 09 program package.[29]

XA spectra were calculated on the first principles multi-reference restricted active space self-consistent field (RASSCF) level with the relativistic ANO-RCC-VTZ basis set for Fe, C, and N atoms.[30-31] This method was recently applied to several transition metal compounds in solution[12, 32-35] The RASSCF calculations implied no symmetry. MOs were first optimized in a state-averaged complete active space calculation for the lowest fifteen valence singlet states and then all occupied MOs except for the active ones were kept frozen. For the L-edge calculations spin-orbit coupling was treated within the state-interaction (RASSI-SO) method[36] including directly interacting singlet and triplet states. Scalar relativistic effects were considered within the Douglas-Kroll-Hess approach.[37-38] RASSCF/RASSI calculations were performed with the MOLCAS 7.8 program suite.[39]

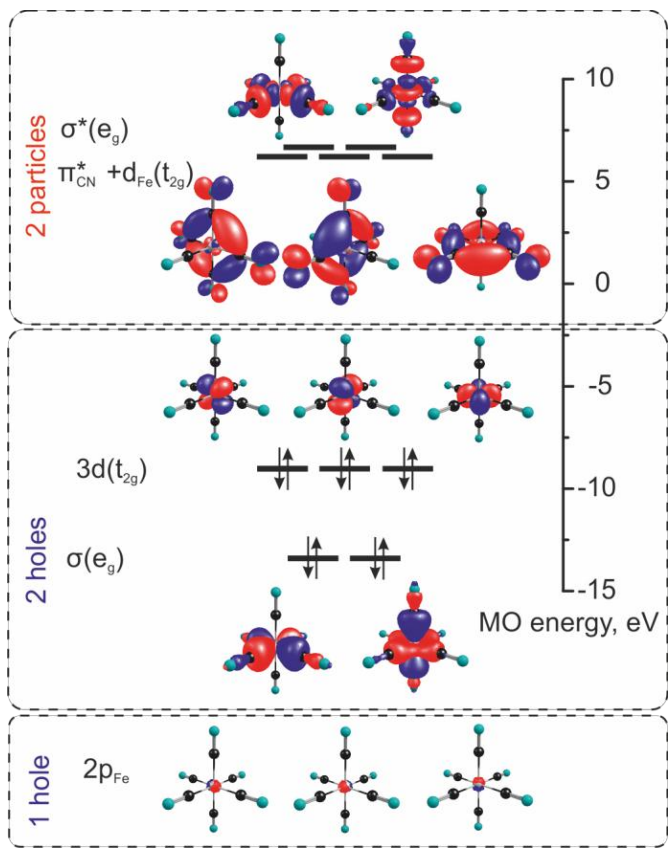

**Figure 1.** The MOs included in the active space for RASSCF calculations of Fe L-edge XAS and RIXS spectra are presented. Their orbital energies were determined by RASSCF PCM ground state calculations. The full list of ligand K-edge active space is presented in supplementary Fig. S1.

For the Fe L-edge XAS and RIXS calculations, the active space comprised 16 electrons distributed over 13 orbitals shown in Figure 1: three $2p_{Fe}$, two σ-bonding $\sigma(e_g)$, three $3d(t_{2g})$, two σ-antibonding $\sigma^*(e_g)$, and three $\pi^*_{CN} + d_{Fe}(t_{2g})$. One hole was allowed for $2p$ orbitals, whereas two holes were allowed for $\sigma(e_g)$ and $3d(t_{2g})$ as well as two particles for $\sigma^*(e_g)$ and $\pi^*_{CN} + d_{Fe}(t_{2g})$. This choice of the active space was dictated by the necessity of the inclusion of correlated d-orbitals to correctly describe static correlation as pointed out in Ref. [40] Allowing for two particle excitations insures the inclusion of the most important correlation terms, still keeping the reasonable number of RASSCF states under consideration. The effect of dynamic correlation was further included via second order perturbation theory (RASPT2) with an imaginary level shift of 0.4 Hartree to avoid intruder states (see supplementary Fig. S3).

In the case of N K-edge calculations, the active space included six $1s_N$ instead of three $2p_{Fe}$ orbitals and was complemented with twelve MOs having mainly bonding $\pi_{CN}$ character to account for relaxation from these orbitals in the RIXS process. Within RASSCF scheme, single (from $1s_N$, $\pi_{CN}$, and $\sigma(e_g)$) and double excitations (from $3d(t_{2g})$) were considered to account for the most important correlation effects (see supplementary Fig. S3).

Note, for the purpose of comparison, the simulated XA spectra were globally shifted to the red with respect to the experiment by 8.5 and 23.7 eV for Fe L-edge and N K-edge, respectively. In general, such a shift can be attributed to insufficient flexibility (minimal basis quality) of the ANO-RCC bases in the core-region and frozen relaxation of a fraction of valence orbitals upon core-hole formation.

RIXS intensities accounting for electronic coherence effects were estimated according to the Kramers-Heisenberg expression;[41] a Gaussian excitation profile (0.5 eV) and Lorentzian lifetime broadening (0.5 eV) were used. The RIXS spectra were additionally convoluted with 1.0 eV Gaussian to mimic experimental linewidths. The effect of polarization dependent detection was included according to Ref.[42]

**RESULTS**

On the left panel of Figure 2 the iron L-edge partial fluorescence yield (PFY) X-ray absorption (black) spectrum of potassium ferrocyanide solution is presented. It is overlaid with results of RASSCF calculations (magenta). Overall, the theoretical spectrum resembles the general shape of the experimental one. A disagreement between theory and experiment is observed for the double peak structures and their intensity ratios both in the $L_3$- and $L_2$-edges, as well as the $L_3/L_2$ spin-orbit splitting. The first discrepancy is related to the lack of dynamic correlation. Upon inclusion of the latter via the RASPT2 approach, the splitting between the subbands decreases, thus improving the agreement with the experiment. However, the agreement in intensity notably deteriorates (see Fig. S3). The double peak intensity ratio is discussed in the light of the d-contributions to the corresponding MOs below. The overall disagreement in the $L_3/L_2$ splitting might be the consequence of the truncation of the basis of RASSCF spin-free functions in the RASSI-SO calculation. Taking into account these discrepancies and their possible sources, the calculations allow for a qualitative assignment of experimental features.

According to the theoretical results, the Fe $L_3$-edge consists of three groups of lines (denoted as 1-3 in Figure 2). The first peak in the PFY spectra at 710.7 eV is due to line 1 attributed to $2p \rightarrow \sigma^*(e_g)$ transitions and line 2 having the same character plus shake up $3d(t_{2g}) \rightarrow \sigma^*(e_g)$ excitations. The second experimental peak at 712.4 eV (corresponding to line 3) is due to mixed excitations from $2p$ to $\pi^*_{CN} + d_{Fe}(t_{2g})$ and $\sigma^*(e_g)$ MOs. Transitions 2 and 3 are to a high extent due to spin-forbidden singlet-triplet excitations gaining intensity through spin-orbit coupling. The peaks at 723.3 eV and 725.2 eV (4 and 5 in Figure 2) are $L_2$-edge replica of the 1+2 and 3 groups of transitions broadened due to shorter core-hole lifetime.

Note that the high intensity of $2p \rightarrow \pi^*_{CN} + d_{Fe}(t_{2g})$ already evidences the strong orbital mixing of ligand $\pi^*_{CN}$ orbitals with metal d-orbitals (see Table 1). In line with the same observation for solid ferrocyanide, this is evidence for π-back-donation.[21] The overall shape and nodal surfaces of the $\pi^*_{CN} + d_{Fe}(t_{2g})$ orbitals (see Figure 1) suggests that these are Rydberg 4d orbitals.

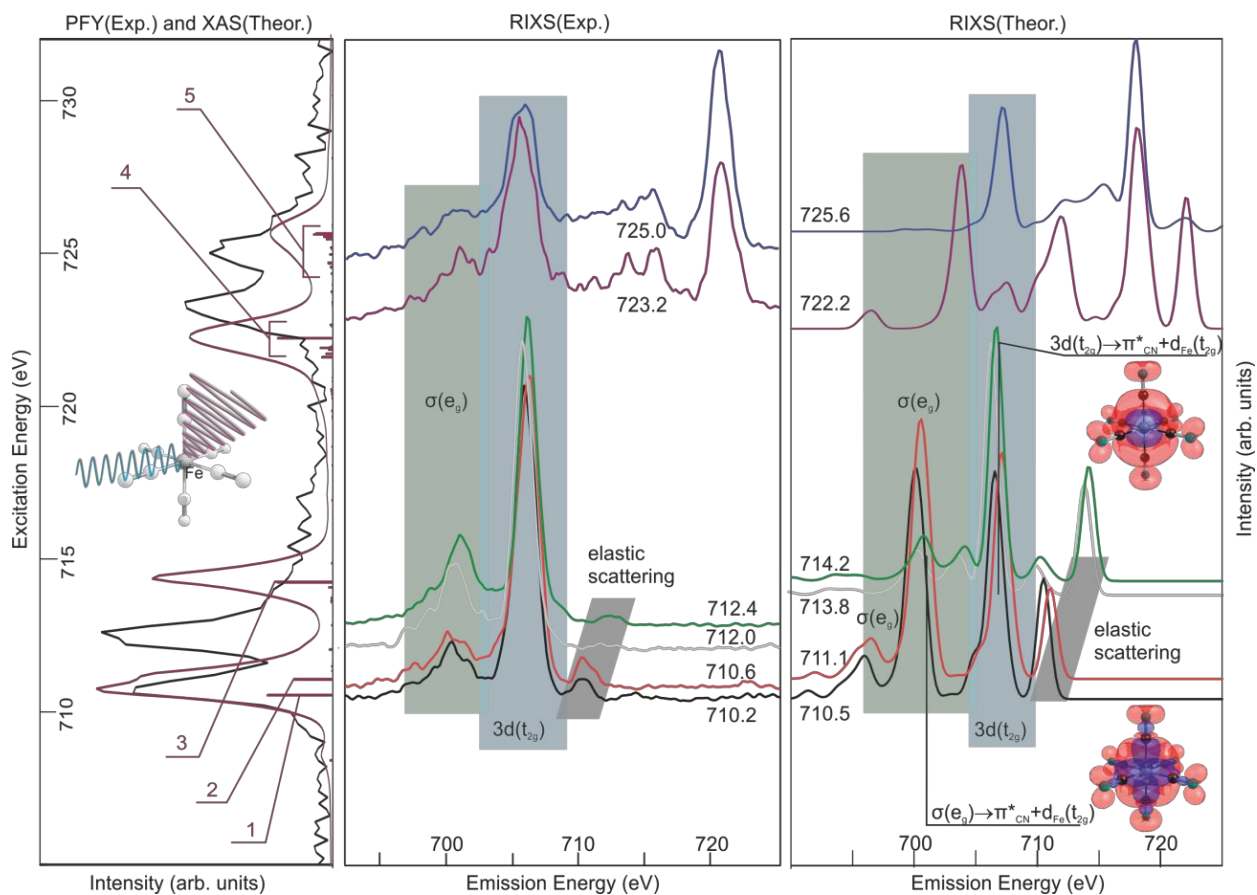

**Figure 2.** Left panel: iron L-edge PFY spectrum (black) as compared to theoretical XAS (magenta). Middle panel: experimental RIXS of potassium ferrocyanide solution. Right panel: theoretical RIXS together with density difference plots showing various types of RIXS relaxation (see text). The red and blue colors denote positive and negative density differences – negative showing localization of the relaxed electron which fills the $2p_{Fe}$ core-hole. For the assignment of transitions labeled with the numbers 1-5, see the text.

The core-excited electronic states are of strongly multi-configurational nature, with singlet and triplet states being additionally strongly mixed due to spin-orbit coupling. Electron correlation is responsible for the reversed order of the core-excited states $\pi^*_{CN} + d_{Fe}(t_{2g})$ and $\sigma^*(e_g)$ as compared to the MO order (Figure 1). Note that, as pointed out by Pierloot et al.,[40] the inclusion of multi-reference effects is important, since correlation leads to notable redistribution of electron density and thus the description of bonding and back bonding. For better clarity, the results are discussed in terms of MOs below. The experimentally determined values for the $L_3/L_2$ spin-orbit splitting (12.6±0.2) eV and the branching ratio (0.58) presented here coincide with those presented elsewhere.[21, 43]

The middle panel of Figure 2 shows the iron L-edge RIXS spectra. Their theoretical counterparts are shown on the right panel (the full 2D RIXS plane is presented in the supplementary Fig. S2). Inclusion of electronic coherence effects changes the RIXS spectra only slightly (Fig. S4). In the middle panel, the gray shaded peaks that shift to higher energy with an increasing excitation energy resemble the elastically scattered X-rays. The theoretical elastic peaks are notably more intense than in the experiment. This might be attributed to the broadening due to the vibrational progression[44] which is not included in the theoretical approach. The theoretical RIXS spectra have different relative intensities for the excitation at the $2p \rightarrow \sigma^*(e_g)$ resonances (first absorption peak in $L_3$ and $L_2$) as compared to the experimental spectra. In this case, the emission bands at about 700.4 eV and 703.4 eV are of similar intensity as the main peak at 706.6-707.8 eV, whereas their intensities are lower in the experimental spectra. In fact, the RASSCF method overestimates the intensity of RIXS transitions corresponding to the $\sigma(e_g) \rightarrow \sigma^*(e_g)$ final states. However, this discrepancy does not prevent from qualitative assignment of the experimental features based on theory. For the $L_3$-edge excitation, the most prominent experimental feature centered at ca. 706.1 eV is assigned to the relaxation from the $3d(t_{2g})$ MOs which are strongly localized at the iron-site. In the right panel, the electronic density difference $\left(\Delta\rho(\vec{r}) = \sum_i \Delta n_i \ |\varphi_i(\vec{r})|^2 \right)$ for the $3d(t_{2g}) \rightarrow \pi^*_{CN} + d_{Fe}(t_{2g})$ RIXS final state is depicted,

whereas $\Delta n_i$ is the change in orbital occupancy upon transition and the summation runs over all active MOs – $\varphi_i(\vec{r})$. The negative density difference of the relaxed electron (blue) which is solely localized on iron is most interesting for the present discussion. The less intense feature centered at ca. 701.0 eV is mainly due to relaxation from $\sigma(e_g)$ bonding MOs which have notable contributions from ligand atoms. However, the quite intense local $2p \leftarrow 3d(t_{2g})$ transitions, mixed with shake up excitations, also contribute to this band. The significant ligand localization of the relaxed electron for $\sigma(e_g) \rightarrow \pi^*_{CN} + d_{Fe}(t_{2g})$ final state ($\sigma(e_g)$ MOs has 23-33% metal 3d character, shown in Table 1) appears as a negative density on the carbon atoms in the density difference plots (Figure 2 right panel). Thus, transitions contributing to this low energy peak are of considerable ligand-to-metal charge-transfer character and demonstrate strong σ-donation in the metal-ligand chemical bond of the ferrocyanide complex.

To probe the orbital mixing strength at the metal-ligand π-bond, the N K-edge XAS and RIXS spectra were measured. The XAS spectrum measured in TFY mode at the K-edge of nitrogen, presented on the left side of Figure 3, has one pronounced broad feature at ca. 400 eV which is assigned to the $1s_N \rightarrow \pi^*_{CN} + d_{Fe}(t_{2g})$ transitions on the basis of RASSCF calculations. The $1s_N \rightarrow \sigma^*(e_g)$ transitions have low intensity because of the small 2p contribution of nitrogen to the $\sigma^*(e_g)$ MO (Table 1) and contribute to the low-energy wing of the absorption band. The peaks at 402.2 eV and 404.6 eV are related to electronic transitions into states of local p-character, i.e., into Rydberg and σ*-type MOs.[45-46] These are of no further relevance for the present study, since the RIXS spectra reveal no changes for excitation energies above the π* resonance.

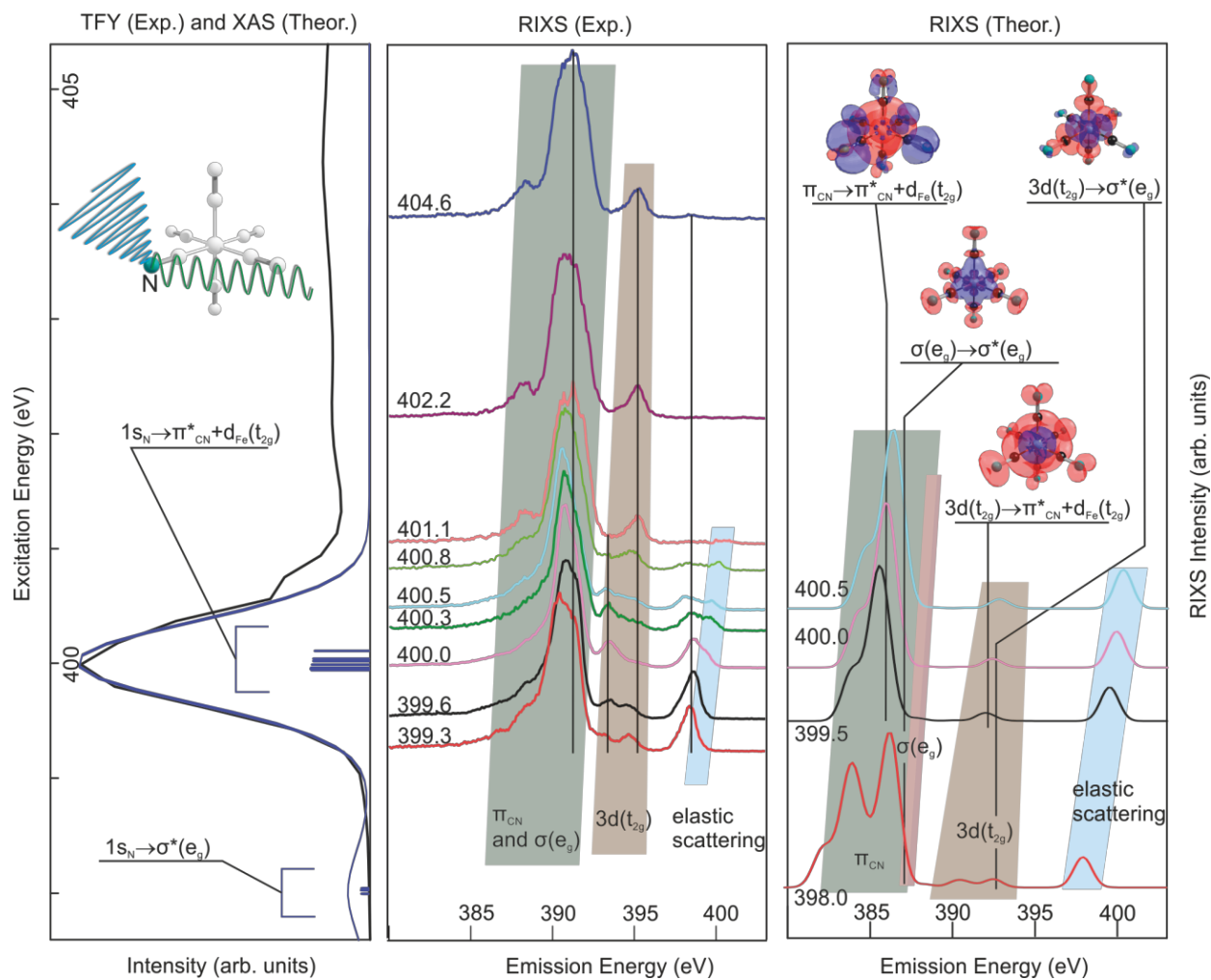

**Figure 3.** Left panel: nitrogen K-edge TFY spectrum (black) as compared to theoretical XAS (blue). Middle and right panels: Experimental and theoretical RIXS of potassium ferrocyanide solution, respectively. The density difference plots show the RIXS relaxation processes of local and charge-transfer types (see text). The red and blue colors denote positive and negative density differences – negative showing localization of the relaxed electron which fills the $1s_N$ core-hole.

The experimental and theoretical RIXS spectra at the nitrogen K-edge are presented in the middle and right panels of Figure 3, respectively. Here, the RIXS features reveal states of local nitrogen p-character according to the (dipole) selection rules. Based on the theoretical assignment, the main peak at about 391 eV is due to relaxation from the 12 MOs having mainly $\pi_{CN}$ character (see Fig. S1) and thus being rather local: the $\pi_{CN} \to \pi_{CN}^* + d_{Fe}(t_{2g})$ negative density difference in Figure 3 is localized almost exclusively on the ligands. The $1s_N \leftarrow \sigma(e_g)$ transitions with notable intensity are also localized in this region and have significant charge-transfer character. The features between the main peak and the elastic band (393-396 eV) are most interesting. They correspond to $1s_N \leftarrow 3d(t_{2g})$ relaxation which, according to density difference plots with metal-localized negative density, can be attributed to metal-to-ligand charge-transfer transitions (only about 1-2% of nitrogen 2p character in $3d(t_{2g})$ MOs, Table 1). The 1s core-excited state $\pi_{CN}^* + d_{Fe}(t_{2g})$ has a strong metal $t_{2g}$ character (55%, see Table 1), thus the strong mixed character of this state gives rise to the pronounced $1s_N \leftarrow 3d(t_{2g})$ transition, and is a manifestation of the strong π-backdonation in the metal-ligand chemical bond of ferrocyanide complex.

**DISCUSSION**

The occurrence of charge-transfer transitions is important evidence of orbital mixing (donation and backdonation). To obtain a qualitative understanding of its appearance in the X-ray spectra we consider a simple model for the cross terms (charge-transfer) and their RIXS intensity. The simplification is that states are single-configurational and the whole process can be well described in terms of selected atomic ($|p_{Fe}\rangle$, $|d_{Fe}\rangle$, $|s_N\rangle$, $|p_N\rangle$, …) and molecular ($|\sigma d\rangle$, $|\pi_{CN}\rangle$, $|\pi_{CN}^* + d_{Fe}\rangle$, …) orbitals. In the case of RIXS at the Fe L-edge, the transition dipole operator is local because of dipole selection rules and strong localization of the $|p_{Fe}\rangle$ orbital,

$$\widehat{D} = d_{p_{Fe} \to d_{Fe}} |d_{Fe}\rangle\langle p_{Fe}| + h.c.,$$

where $d_{p_{Fe} \to d_{Fe}}$ is a number characterizing the transition intensity.

For resonant excitation to the $|\pi^*_{CN} + d_{Fe}\rangle$ state and emission from the same state and from the occupied $|\sigma d\rangle$ state, the wave functions directly after absorption (t = 0) and after emission (t = τ) are

$$|\Psi(t=0)\rangle = \hat{D}|p_{Fe}\rangle = d_{p_{Fe} \to d_{Fe}} C_d^{\pi^*_{CN}+d_{Fe}}(t=0)\,|d_{Fe}\rangle,$$

$$\langle\Psi^{\pi^*_{CN}+d_{Fe}}(t=\tau)| = \langle\pi^*_{CN}+d_{Fe}|\hat{D} = d_{p_{Fe}\to d_{Fe}} C_d^{\pi^*_{CN}+d_{Fe}}(t=\tau)\langle p_{Fe}|, \text{ and}$$

$$\langle\Psi^{\sigma d}(t=\tau)| = \langle\sigma d|\hat{D} = d_{p_{Fe}\to d_{Fe}} C_d^{\sigma d}(t=\tau)\langle p_{Fe}|,$$

for relaxation from $|\pi^*_{CN}+d_{Fe}\rangle$ and $|\sigma d\rangle$ states, respectively; τ is the duration of the scattering process. The coefficients $C_d(t)$ are the contributions of the corresponding atomic orbitals (in this case of d-character) to the MOs. For the completely relaxed case (t=∞) the corresponding coefficients as predicted by the RASSCF method are given in Table 1 (values in parentheses). The intensity of a RIXS band according to the Kramers-Heisenberg expression[41] is proportional to

$$|\langle p_{Fe}|\Psi(t=0)\rangle\langle\Psi(t=\tau)|p_{Fe}\rangle|^2,$$

what gives

$$(d_{p_{Fe}\to d_{Fe}})^4 (C_d^{\pi^*_{CN}+d_{Fe}}(t=0) C_d^{\pi^*_{CN}+d_{Fe}}(t=\tau))^2$$

and

$$(d_{p_{Fe}\to d_{Fe}})^4 (C_d^{\pi^*_{CN}+d_{Fe}}(t=0) C_d^{\sigma d}(t=\tau))^2$$

for relaxation from $|\pi^*_{CN}+d_{Fe}\rangle$ and $|\sigma d\rangle$ states. Thus the ratio of intensities of two RIXS bands, one of them being a charge-transfer band and the other one a local excitation band, in simplified form can be a measure of the ratio $C_{3d}^{\sigma d}(t=\tau)/C_{3d}^{\pi^*_{CN}+d_{Fe}}(t=\tau)$. The same holds for the nitrogen K-edge, where again, e.g., $C_{p_N}^{dt_{2g}}(t=\tau)/C_{p_N}^{\pi_{CN}}(t=\tau)$ can be considered as a measure of the $2p_N$ contribution to the $3d(t_{2g})$ MO.

Summarizing our qualitative model, the locality of X-ray excitation operators allows projecting the MOs onto atomic orbitals which is not possible for, e.g., UV/Vis spectroscopy in such selective way. In addition, XAS can only probe contributions to unoccupied states, whereas RIXS allows for complementary probing of the composition of occupied states.

The σ- and π-donation/backdonation are straightforwardly interpreted in VB theory, e.g., ligand-field multiplet theory as applied in ref. [21] However, in MO type calculations, the orbital mixing concept allows for a more direct interpretation. Here, orbital coefficients are directly related to the intensities of XAS and RIXS transitions through the dipole selection rules as described above. In the present work we measured orbital mixing in four groups of orbitals and thus three situations illustrate bonding and back-bonding:

1. The mixing between $\pi_{CN}^*(t_{2g})$ and $3d(t_{2g})$ in unoccupied MOs (π-back-donation) determines the intensities of the $2p \rightarrow \pi_{CN}^* + d_{Fe}(t_{2g})$ iron L-edge XAS absorption bands.

2. Analogously the d-contribution in $\sigma^*(e_g)$ orbitals determines the intensity of $2p \rightarrow \sigma^*(e_g)$ XAS bands.

3. The relative intensities of metal-to-ligand charge-transfer and elastic peaks in nitrogen K-edge RIXS again reflects mixing between $3d(t_{2g})$ and $\pi_{CN}^*(t_{2g})$ MOs but now for the occupied counterpart.

4. The relative intensities of ligand-to-metal charge-transfer and elastic bands in iron L-edge RIXS determine the mixing between $\sigma_{CN}$ and iron $3d(e_g)$ MOs and thus σ-donation strength.

In the following, we compare these results with those from our previous study of the Fe(CO)$_5$ complex in order to draw conclusions on the orbital mixing and donation/backdonation properties of these two iron complexes. Note that for Fe(CO)$_5$ no solvent was included in theoretical calculation. In [Fe(CN)$_6$]$^{4-}$ iron L-edge XAS (Figure 2), the intensities of $2p \to \sigma^*(e_g)$ and $2p \to \pi^*_{CN} + d_{Fe}(t_{2g})$ are almost equal, whereas in Fe(CO)$_5$ the second transition is much more intense than the first one.[12] Since the strength of π-back-donation depends on the strength of σ-donation in XAS, the relative intensities of these two peaks must be considered. This shows that in Fe(CO)$_5$ the CO ligand is a stronger π-acceptor than CN$^-$ in [Fe(CN)$_6$]$^{4-}$. The same conclusion follows from the comparison of the charge-transfer vs. elastic peak intensities in nitrogen (for ferrocynanide) and oxygen (for iron pentacarbonyl[12]) K-edge RIXS spectra, whereas for the latter the ratio is ~2.4 times larger. In contrast, the charge-transfer vs. elastic intensity ratio in the iron L-edge RIXS is ~2.2 times larger, thus revealing CN$^-$ as a stronger σ-donor than CO. Whereas the conclusion itself is obvious due to the negative charge of CN$^-$, here the quantification of the strength is experimentally achieved using experimental RIXS spectra.

These conclusions can also be illustrated by the atomic orbital contributions of d$_{Fe}$ and p$_N$ character to the MOs (Table 1) which are relevant to the discussion according to the dipole-selection rules. Note that the total d-character of σ* and π* orbitals of 236% from our data agrees well with the value of 287(30)% from ref.[21] In Comparison to the ground state DFT calculations from the same reference, the RASSCF method predicts similar σ-donation effect (33% of d-character in $\sigma(e_g)$ MOs). However, in case of RASSCF, the d-contributions to $3d(t_{2g})$ and $\pi^*_{CN}(t_{2g})$ are notably larger (99% (RASSCF) vs. 77% (DFT) and 55%(RASSCF) vs. 16% (DFT), respectively) and to $\sigma^*(e_g)$ smaller (35% (RASSCF) vs. 57% (DFT)). Note that orbital coefficients substantially differ from the ground state values if the relaxation upon core-hole formation is taken into account (Table 1, values in parentheses). The most pronounced change is in the composition of ligand π*orbitals and orbital relaxation decreases their d-character. In fact the intensities of XAS bands depend on $C(t=0)$ and RIXS intensities depend on $C(t=\tau)$, where τ is the duration of a scattering process. Values in Table 1 correspond to $C(t=0)$ (ground state) and $C(t=\infty)$ totally relaxed intermediate core-excited state. This means that RIXS measures orbital composition between these two limiting cases. Noteworthy, though RASSCF predicts larger orbital mixing for $\pi^*_{CN}(t_{2g})$ orbital than for $\sigma^*(e_g)$ the total intensity of the corresponding absorption band is underestimated. Note that inclusion of dynamic correlation (RASPT2) even lowers the intensity of $\pi^*_{CN}(t_{2g})$ band (Fig. S3).

| d-character[a] | [Fe(CN)$_6$]$^{4-}$ | | | | Fe(CO)$_5$[b] | | | |
|---|---|---|---|---|---|---|---|---|
| | MO | $d_{Fe}$ | $p_N$ | Occ.[c] | MO | $d_{Fe}$ | $p_O$ | Occ.[c] |
| $d_{z^2}$ | $\sigma(e_g)$ | 33.1(22.8) | 3.5(4.1) | 1.994 | $\sigma(a'_1)$ | 15.9(12.6) | 3.1(3.4) | 1.985 |
| $d_{x^2-y^2}$ | | 33.1(22.9) | 3.8(4.3) | 1.994 | $3d_\sigma(e')$ | 86.7(76.0) | 5.7(4.2) | 1.976 |
| $d_{xy}$ | $3d(t_{2g})$ | 98.5(96.8) | 0.6(2.1) | 1.975 | | 86.7(76.3) | 5.7(4.2) | 1.976 |
| $d_{xz}$ | | 98.5(96.8) | 0.6(2.1) | 1.972 | $3d_\pi(e'')$ | 91.4(90.4) | 6.3(7.3) | 1.980 |
| $d_{yz}$ | | 98.5(96.8) | 0.8(2.7) | 1.972 | | 91.4(90.4) | 6.3(7.3) | 1.980 |
| $d_{z^2}$ | $\sigma^*(e_g)$ | 34.9(32.1) | 2.7(2.7) | 0.010 | $\sigma^*(a'_1)$ | 36.6(30.8) | 1.5(1.3) | 0.025 |
| $d_{x^2-y^2}$ | | 34.9(32.0) | 3.6(3.9) | 0.011 | $\pi^*_{CO}(e')$ | 34.6(14.7) | 20.8(29.8) | 0.021 |
| $d_{xy}$ | $\pi^*_{CN}(t_{2g})$ | 56.1(5.4) | 37.2(37.4) | 0.023 | | 34.6(9.6) | 20.8(19.5) | 0.021 |
| $d_{xz}$ | | 55.0(5.3) | 25.3(39.6) | 0.025 | $\pi^*_{CO}(e'')$ | 36.4(7.9) | 26.0(16.1) | 0.019 |
| $d_{yz}$ | | 55.0(5.3) | 29.0(47.6) | 0.025 | | 36.4(8.1) | 26.0(16.6) | 0.019 |

Table 1. Contributions of atomic orbitals of d$_{Fe}$ and p$_N$ character to the MOs of [Fe(CN)$_6$]$^{4-}$ and Fe(CO)$_5$ molecules in the ground state. The values in parentheses correspond to orbitals relaxed

upon core-hole formation. Due to small asymmetry, the contributions to degenerate MOs are slightly different. [a] Symmetry of the d-orbital which mainly forms the MO; MOs of $[Fe(CN)_6]^{4-}$ and $Fe(CO)_5$ are arranged accordingly. [b] See ref.[12] [c] Occupancies of the corresponding MOs in the ground electronic state.

If compared to $Fe(CO)_5$, the d-contribution to the $\sigma(e_g)$ MO is twice larger for ferrocyanide, reflecting stronger σ-donation as discussed above. The d-contributions to $\pi^* + d_{Fe}$ orbitals are smaller for $Fe(CO)_5$ in the ground state and larger for core-excited states but they should be considered relative to σ-donation strength for both compounds. Note that apart from the one-electron donation/back-donation effect (orbital mixing), the density flux due to many-electron effects (electron correlation) can be essential.[40] In Table 1, the occupancies of orbitals in the RASSCF multi-configurational wave function are presented. These occupancies show that correlation effects are comparable for both compounds and main density redistribution effects are due to orbital mixing.

Direct measurements of orbital mixing for different groups of MOs combining metal L-edge XAS and both ligand K-edge and metal L-edge RIXS can provide essential information for the TM coordination chemistry and in particular to the understanding of catalytic activity of TM complexes. In contrast to structural investigations (bond lengths, vibrational force constants) and atom-unspecific spectroscopic methods (e.g., UV/VIS) where the effects of bonding and back-bonding can hardly be decoupled from each other, the combination of X-ray spectroscopic techniques allow distinguishing between these effects due to projection of MOs onto atomic contributions of particular angular momentum.

**CONCLUSIONS**

In the present article we addressed the nature of the chemical bond in potassium ferrocyanide in aqueous solution using X-ray spectroscopic techniques together with high-level first principles calculations. Our results agree well with previous publications, but go beyond them, especially due to the recent experimental development, allowing for the application of soft X-ray spectroscopy on liquid phase samples as well as the employment of state-of-the-art theoretical methods.

Whereas XAS was used for probing the unoccupied MOs, RIXS was employed for concluding on the nature of occupied MOs. Relative intensities of charge-transfer bands with respect to local transitions were related to the coefficients of atomic orbitals in the corresponding MOs and thus RIXS spectra allowed to conclude on the strength of orbital mixing between metal and ligands. Using soft X-ray absorption and resonant inelastic X-ray scattering at both central metal L-edge and ligand nitrogen K-edge, direct experimental observation of the local electronic structure as well as σ-donation and π-backdonation in the TM complex was presented. Strong ligand-to-metal charge-transfer as evidence for strong σ-donation and strong metal-to-ligand contributions showing π-backdonation were observed. Comparison to data from ref.,[12] shows that, in the case of the aqueous ferrocyanide ion, σ-donation is stronger and π-back-donation weaker than in case of iron pentacarbonyl.

The obtained information presents a step towards understanding the origin of the properties of TM complexes as well as their functioning in industrial and biological catalytic reactions, based on the special nature of their chemical bonds, since these are closely related to ligand exchange and thus bond formation and cleavage. Accordingly, with this study, we hope to motivate future studies of complex catalysts in solution especially using the powerful combination of XAS,

RIXS, and advanced theoretical calculations to shed further light on the peculiarities of such species.

## ABBREVIATIONS

TM, transition metal; XAS, X-ray absorption spectroscopy; RIXS, resonant inelastic X-ray scattering; MO, molecular orbital; VB, valence bond theory; $[Fe(CN)_6]^{4-}$, ferrocyanide, hexacyanoferrate(II); $CN^-$, cyanide; TFY, Total fluorescence yield; PFY, partial fluorescence yield; RASSCF, restricted active space self consistent field; RASSI-SO, restricted active space state interaction with spin-orbit coupling.

## AUTHOR INFORMATION

### Corresponding Authors

*Emad F. Aziz, email:  emad.aziz@helmholtz-berlin.de

*Sergey I. Bokarev, email:   sergey.bokarev@uni-rostock.de

### Notes

The authors declare no competing financial interest.

### Author Contributions

The experiments were performed by N.E., R.G.D., E.S., K.M.L., R.G., M.D., A.K., and K.A, while the theoretical calculations were performed by S.I.B. N.E. and E.S. evaluated the data. N.E., S.I.B., E.S., O.K., and E.F.A. analyzed the data and discussed the results. N.E., S.I.B., and E.S. wrote the manuscript with the help of all authors' contributions. E.F.A. designed and E.F.A. and O.K. coordinated the project. All authors have given approval to the final version of the manuscript.

## ACKNOWLEDGMENT

This work was supported by the Helmholtz-Gemeinschaft young investigator fund VH-NG-635 and the European Research Council grant No. 279344. K.A. would like to acknowledge the financial support of the Einstein Foundation Berlin for the postdoctoral scholarship.

## SUPPORTING INFORMATION AVAILABLE

Active space for the nitrogen K-edge XAS and RIXS RASSCF calculations.

This information is available free of charge via the Internet at http://pubs.acs.org

# Chemical Bonding in Aqueous Ferrocyanide: Experimental and Theoretical X-ray Spectroscopic Study

Nicholas Engel,[1] Sergey I. Bokarev,*,[2] Edlira Suljoti,[1] Raül Garcia-Diez,[1] Kathrin M. Lange,[1] Kaan Atak,[1] Ronny Golnak,[1] Alexander Kothe,[1] Marcus Dantz,[1] Oliver Kühn,[2] Emad F. Aziz,*,[1]

**Supplemetary Material**

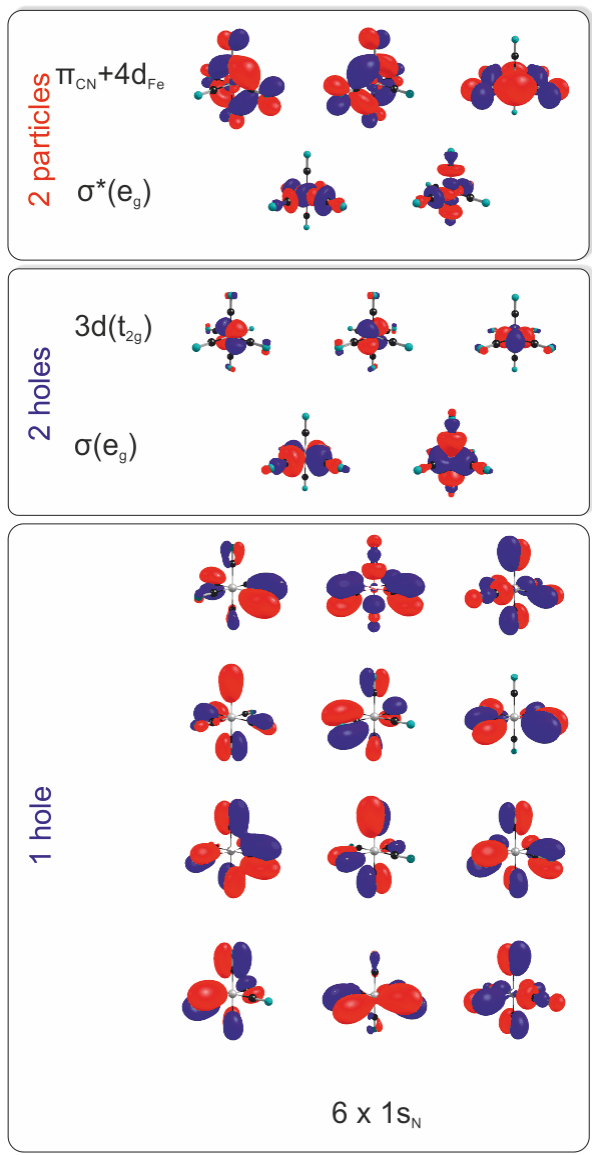

Figure S1. Orbital subspaces used for RASSCF/RASSI calculation on N K-edge XAS and RIXS spectra.

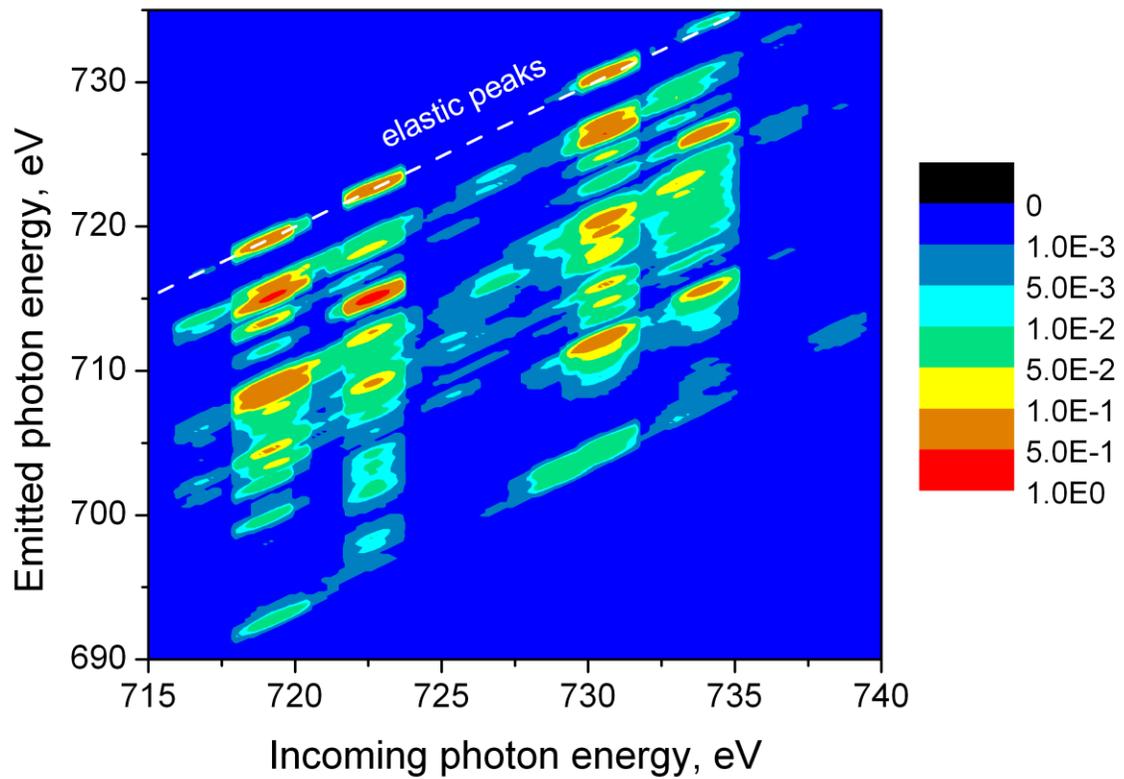

Figure S2. Normalized 2D RIXS plot of iron L-edge of [Fe(CN)$_6$]$^{4-}$ as predicted by RASSCF(PCM).

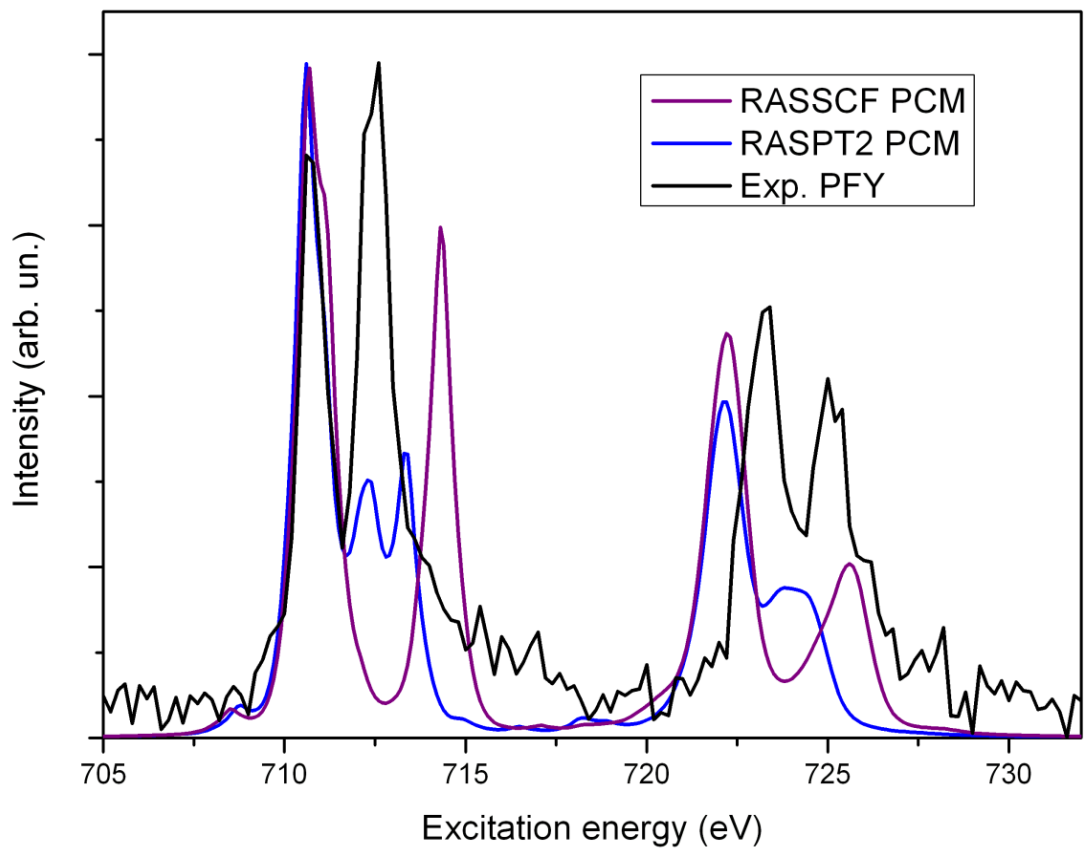

Figure S3. Comparison of RASSCF and RASPT2 XAS with experimental PFY spectrum. Pseudo-Voigt lineshape is used: $L_3$-edge 0.5L(0.3)+0.5G(0.9) and $L_2$-edge 0.5L(0.75)+0.5G(0.9), where L and G stand for Lorentzian and Gaussian shapes, widths are in eV.

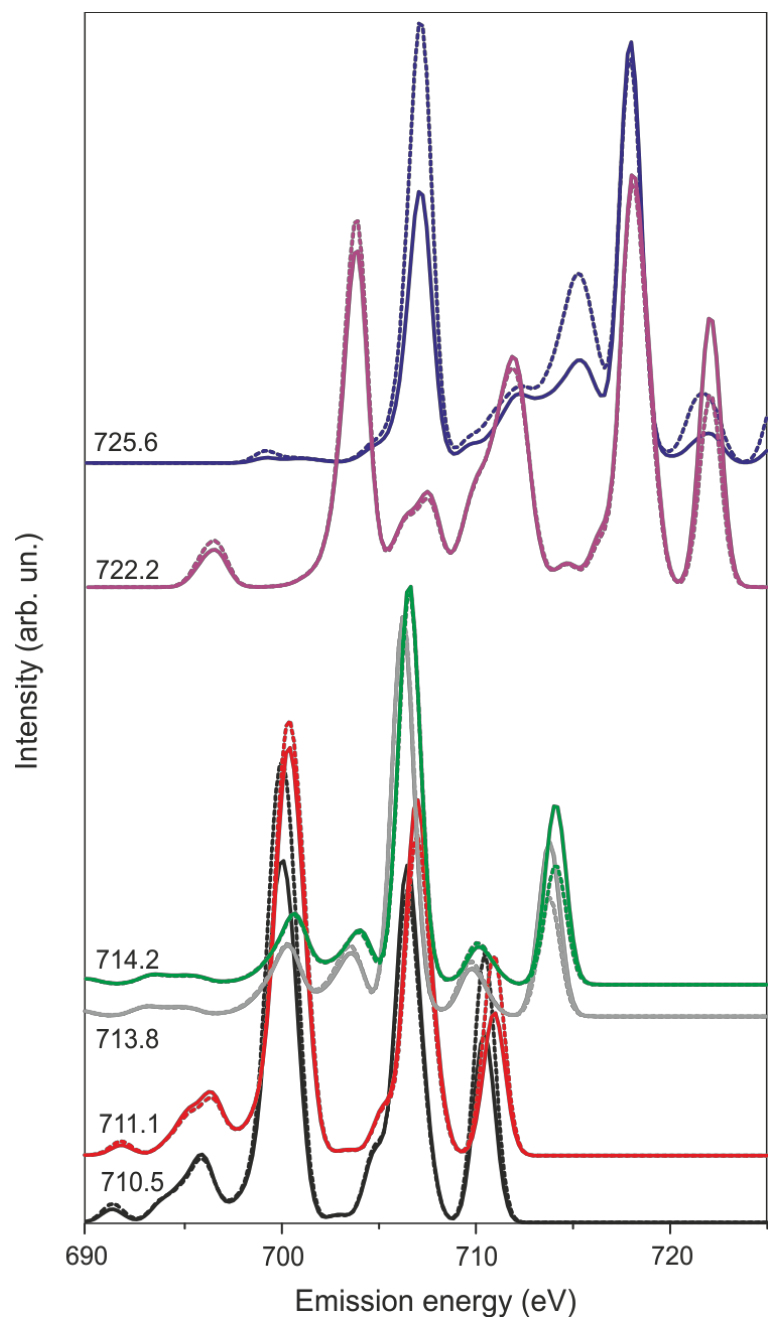

Figure S4. Theoretical iron L-edge RIXS spectra with account for electronic coherences (solid lines) and without (dashed lines).